# Towards enterprise-ready AI deployments

Minimizing the risk of consuming AI models in business applications

Vinod Muthusamy
IBM Research AI

Aleksander Slominski
IBM T.J. Watson Research Center

Vatche Ishakian
Bentley University

*Abstract*—The stochastic nature of artificial intelligence (AI) models introduces risk to business applications that use AI models without careful consideration. This paper offers an approach to use AI techniques to gain insights on the usage of the AI models and control how they are deployed to a production application.

## I. INTRODUCTION

Artificial intelligence (AI), including deep learning, have revolutionized business applications in diverse fields, including finance, manufacturing, and logistics. For businesses, adopting AI presents an opportunity and a risk. On the one hand, AI can reduce cost or provide better customer experience, with a $15.7 trillion "potential contribution to the global economy by 2030" [1]. On the other hand, adopting AI models present risks that can manifest in the form of monetary or reputation loss. For example, data used for training may be manipulated [2] or contain implicit racial, gender, or ideological biases [3].

This paper proposes an architecture that allows a set of AI algorithms to be transparently plugged into a business application. An application owner can understand how the AI models are used in the context of a business application and apply techniques to continuously but carefully promote new models to production without breaking the application.

## II. MODELS AS MICROSERVICES

Consider an AI model with a scoring function that's exposed as an API. A business application can invoke this API to make a prediction and recommend the next best action or offer insights about the process to a case worker.

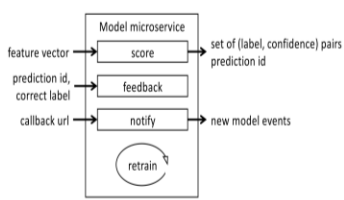

*Figure 1: Model microservice*

We propose that models be treated as microservices [4] with three well-defined interfaces, as illustrated in Figure 1. First, the **score** interface is used to perform scoring on a model. It takes a feature vector as input, and outputs a set of (result, uncertainty) pairs. Second, the **feedback** interface is used to flag good and bad predictions. It's an opportunity to offer labeled data to retrain a model. Finally, the **notify** interface is used to receive alerts when a new version of a model is available. These interfaces of a *model microservice* make it possible to plug in other capabilities, some of which are outlined in the next section.

Continuous training, or the ability to incrementally retrain models with new data, is an important enterprise capability [1]. In our architecture, it is the responsibility of the model microservice to decide when and how to retrain itself, either based on the feedback data, new training data, or a change to the model definition. Every change to the model, however, results in a new version of the model, and it is up to the client to determine which version of the model to use. Incidentally, this makes a serverless [5] inferencing runtime a good fit, both in terms of the performance and cost of having many versions of models ready to be used.

There are well-known practices on how to safely deploy a microservice, including staged deployment techniques such as canary deployments and A/B testing [4]. These can be used just as well when the microservices happen to be models.

Models microservices, however, are different from generic microservices in that their outputs have uncertainty. Furthermore, the implementation of a model is generated not authored by a developer, making it difficult to reason about correctness or corner cases, or develop comprehensive test suites. These differences make models more unpredictable and hence risky to use in a business application. The following section offers techniques to consume these models in a more controlled manner.

## III. CONSUMING MODELS IN BUSINESS APPLICATIONS

Consider the case in Figure 2, where a new version of a model is available and the business application is reconfigured to use it. Even if the new model has been well-tested, it is difficult to foresee the effect on the overall business application.

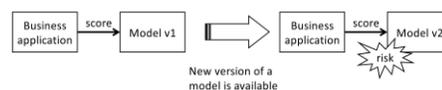

*Figure 2: New version of a model may introduce risks*

A principled approach to consuming AI models can minimize any negative effects of a new model on a business application. This section outlines several techniques to achieve this, built on top of the model microservice definition in Section II. Central to these techniques is the model proxy, which brokers interactions between the business application and the model microservices, as depicted in Figure 3.

The model proxy has the same interface as a model microservice and hence behaves as a transparent proxy. It logs all the calls to the actual models, including the input feature vectors and output results of the **score** interface. Analytics algorithms can be applied to these logs to understand how the models are being used, including execution time metrics, input feature clusters, and uncertainty trends in the model outputs.

These statistics, in turn, can be used to route requests to the appropriate model.

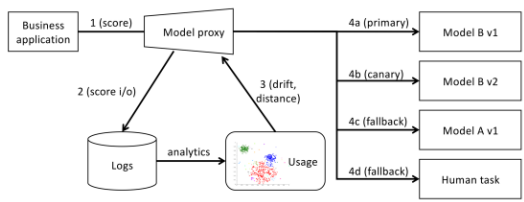

Figure 3: Context-aware proxying to models

Note that the model proxy is both application and model agnostic and can be transparently plugged in. It can, however, be configured with business policies on how new models should be promoted and fallback logic when anomalous behavior is detected.

### A. Usage dashboard

Since the structure of the logs and usage statistics are model agnostic, it is possible to automatically create dashboards showing how models have behaved and been used. For example, the distribution of output labels and confidences over time can be plotted. These dashboards can be used to diagnose errors in the model, or how the model is used by the business process. For example, input features can be clustered using techniques such as k-means [6] in order to detect anomalous inputs or determine which input clusters lead to bad business outcomes.

To make it easy for business users to understand, one approach could be to have a short note, such as the "fact box" depicted in Figure 4, that describes key facts about the deployed AI model version and understand how it compares to previous versions.

Figure 4: Facts about deployed AI model

### B. Data drift detection

Since input feature vectors are logged, algorithms to detect data drift [6] can be used with no additional configuration. There could be both aggregate and individual measures of drift, and anomalous inputs could be defined as those that don't belong to historical input clusters. Drift analysis is useful to data scientists to determine how and when to retrain models, and to application owners to understand how customer behavior is changing.

### C. Staged deployment

A new version of a model can be deployed as a canary, a well-known technique whereby a small percentage of the traffic is routed to the new version [4]. With model microservices, we can build more sophisticated techniques to stage the use of a new model to an application.

*Thresholding* exploits the statistical nature of a model's output so that only high confidence predictions of a new model are used. In Figure 3, the result of invocation 4b determines whether the new model is used. Otherwise, the older model (invocation 4a) is used instead. The confidence thresholds can start high, and lowered as we observe the effect of the new model on the business key performance indicators (KPIs).

*Input clustering* can be used to only use a new model for data that looks similar to input data in the past. Well known algorithms can be used to compute the distance of the live data with historical clusters [6]. An extension of this is to also consider the output clusters, only using the new model for clusters that have delivered high confidence or good business KPIs in the past.

*Exception handling* policies can be defined in cases where predictions have low confidence or exhibit drift. In these cases, the model proxy can fallback to using a model that may not have as good accuracy but whose behavior is well understood (invocation 4c in Figure 3) or revert to asking a case worker to step in (invocation 4d). The ability to fall back to a human decision is a useful property of many business applications.

### D. Continuous training

The model proxy is aware of new versions of the model (via the notify callbacks), and the above staged deployment techniques can be used to automatically rollout new versions of a model in a controlled manner. In addition, multiple models can be run side-by-side, with differences in the outputs flagged for audit or for the data scientist to investigate. This offers a way to test and improve new models in production.

## IV. CONCLUSIONS

The effect of AI models on business applications can be unpredictable due to the stochastic nature of many machine learning models. This paper proposes a data-driven approach to apply AI techniques to provide a high-level view of the risk of using AI in an enterprise application by understanding how models are being used and the effect of models on business KPIs. Business owners are able to make decisions about deploying AI and track its performance over time with the assurance that they get early warning about unexpected behavior and quickly, and in many cases automatically, roll back to previous versions. Ultimately these techniques help reduce the risk of AI models adversely affecting a business.